\documentclass[letterpaper, 10 pt,conference]{ieeeconf}  
\usepackage{amssymb}
\usepackage{amsmath,mathrsfs}
\usepackage{color}
\usepackage{graphicx}
\usepackage{mathtools}
\usepackage{lineno,hyperref}
\usepackage{cleveref}
\usepackage{bm}
\let\labelindent\relax
\usepackage{enumitem}
\usepackage{tikz}
\usepackage{algorithm}
\usepackage{algorithmicx}
\usepackage{algpseudocode}
\modulolinenumbers[5]
\newtheorem{theorem}{Theorem}[section]
\newtheorem{remark}{Remark}[section]

\newtheorem{assumption}{Assumption}[section]

\newtheorem{corollary}{Corollary}[section]

\newtheorem{proposition}[theorem]{Proposition}

\newtheorem{definition}{Definition}[section]

\usepackage{multirow}
\usepackage{color}
\usepackage{cite}

\usepackage{eso-pic}
\AddToShipoutPictureBG*{%
	\AtPageUpperLeft{%
		\setlength\unitlength{1in}%
		\hspace*{\dimexpr0.5\paperwidth\relax}
		\makebox(0,-1.75)[c]{
			\begin{tabular}{c c}
				Rodrigo A. Gonz\'alez \emph{et al.},
				On the Relation between Discrete and Continuous-time Refined Instrumental Variable Methods. \\
				To appear in
				{\em IEEE Control Systems Letters (L-CSS)},
				2023,
				uploaded to arXiv on May 31st, 2023 \\
\end{tabular}}}}

\providecommand{\keywords}[1]{\textbf{\textit{Index terms---}}}

\IEEEoverridecommandlockouts                              
\title{\LARGE \bf On the Relation between Discrete and Continuous-time Refined Instrumental Variable Methods}

\author{Rodrigo A. Gonz\'alez, Cristian R. Rojas, Siqi Pan, James S. Welsh 
\thanks{This work was partly supported by the Netherlands Organization for Scientific Research (NWO) under the research program VIDI with project number 15698, by the Digital Futures project EXTREMUM, and by the Swedish Research Council under contract number 2016-06079 (NewLEADS). R. A. Gonz\'alez is with the Control Systems Technology Group, Eindhoven University of Technology, The Netherlands. C. R. Rojas is with the Division of Decision and Control Systems, KTH Royal Institute of Technology, Stockholm, Sweden. S. Pan and J. S. Welsh are with the School of Engineering, University of Newcastle, Callaghan, Australia. E-mail: r.a.gonzalez@tue.nl.}%
}
\begin{document}

\maketitle


\begin{abstract}
	The Refined Instrumental Variable method for discrete-time systems (RIV) and its variant for continuous-time systems (RIVC) are popular methods for the identification of linear systems in open-loop. The continuous-time equivalent of the transfer function estimate given by the RIV method is commonly used as an initialization point for the RIVC estimator. In this paper, we prove that these estimators share the same converging points for finite sample size when the continuous-time model has relative degree zero or one. This relation does not hold for higher relative degrees. Then, we propose a modification of the RIV method whose continuous-time equivalent is equal to the RIVC estimator for any non-negative relative degree. The implications of the theoretical results are illustrated via a simulation example.
\end{abstract}

\section{Introduction}
System identification deals with the problem of obtaining mathematical models of dynamical systems from data \cite{ljung1998system}. A distinction is made between discrete-time (DT) and continuous-time (CT) models. In DT system identification, it is assumed that a complete description of the system can be made by only observing its behavior at specific time
instants. On the contrary, CT system identification seeks models that reflect the properties of the system for any moment in time. 

There are two main approaches to CT system identification: indirect and direct \cite{garnier2008book}. The indirect approach consists in estimating a DT model with the data, and then converting this model into continuous-time. On the other hand, the direct approach does not use intermediate DT models. It is known that the standard indirect approach leads to models with relative degree one, independent of that of the strictly proper CT system \cite{gonzalez2018asymptotically}; such issue does not arise in the direct approach, since any number of zeros can be accommodated without the need of an additional optimization~step.

For either the direct or indirect approach, refined instrumental variable methods can be applied. The Refined Instrumental Variable (RIV) method for DT systems and its simplified embodiment (SRIV, \cite{young1976some}) are used for estimating Box-Jenkins and output error models respectively, and the CT equivalents of their estimated models are commonly used for initializing the direct approach \cite{garnier2015direct,ljung2009experiments}. Some of the most celebrated direct methods are the CT variants of these estimators, called RIVC and SRIVC \cite{young1980refined}. The RIVC and SRIVC algorithms compute iterative instrumental variable steps by prefiltering the data through CT filters, and are initialized by the CT equivalent of the RIV or SRIV estimators in the MATLAB Contsid Toolbox \cite{garnier2021new}. 

The RIV methods have been used for modeling climate dynamics \cite{young2012recursive} as well as mechanical systems \cite{pan2021continuous}, and extensions of these estimators have been proposed for LPV \cite{laurain2010refined}, Hammerstein-Wiener \cite{ni2013refined} and unstable systems \cite{gonzalez2022refined}. A comprehensive overview of these methods can be found in \cite{young2015refined}; however, such work overlooks a relation between the converging points of the DT and CT RIV~variants.

In this paper, we show that the indirect approach with the RIV estimator provides the same estimate as RIVC at convergence of its iterations (considering \textit{finite} sample size) for CT systems with relative degree one or zero. Although this result no longer holds for higher relative degrees due to parsimony issues, we propose a modification to the RIV estimator, termed Adapted RIV (ARIV), that is shown to extend this equivalence for any non-negative relative degree. As a byproduct, we show that the ARIV estimator can impose relative degree constraints in its CT equivalent without additional optimization steps. The theoretical results include the relationship between the simplified versions of these estimators (SRIV and SRIVC) as a special case.

The remainder of the paper is organized as follows. In Section \ref{context}, we describe the system and model. Section \ref{sec:RIV} presents the notation of the unified RIV estimator; this estimator is later analyzed in Section \ref{sec:equivalence}, which contains the main contributions. In Section \ref{simulations} we provide a simulation example, and we conclude the paper in Section \ref{conclusions}.

\section{System and model setup}
\label{context}
Consider the single-input single-output, linear and time-invariant (LTI), asymptotically stable, CT system 
\begin{equation}
	x(t) =  \frac{B_\textnormal{c}^*(p)}{A_\textnormal{c}^*(p)} u(t), \notag
\end{equation}
where $p$ is the Heaviside operator (i.e., $pu(t)=\textnormal{d}u(t)/\textnormal{d}t$), and $u(t)$ is the input. The numerator and denominator polynomials $B_\textnormal{c}^*(p)$ and $A_\textnormal{c}^*(p)$ are coprime and given~by
\begin{align}
	A_\textnormal{c}^*(p)&= a_n^*p^n+a_{n-1}^*p^{n-1}+\cdots + a_1^*p + 1, \notag \\
	B_\textnormal{c}^*(p)&= b_m^*p^{m}+b_{m-1}^*p^{m-1}+\cdots + b_1^*p + b_0^*, \notag
\end{align}
with $a_n^*\neq 0$, and $n\geq m$. These polynomials are jointly described by the parameter vector 
\begin{equation}
	\label{ctparametervector}
	\bm{\theta}_{\textnormal{c}}^* = \begin{bmatrix}
		a_1^*, & \hspace{-0.15cm}a_2^*, & \hspace{-0.15cm}\dots, & \hspace{-0.15cm}a_n^*, & \hspace{-0.15cm}b_0^*, & \hspace{-0.15cm}b_1^*, & \hspace{-0.15cm}\dots, & \hspace{-0.15cm}b_m^*
	\end{bmatrix}^\top.
\end{equation}
A noisy output measurement is retrieved every $h$[s], i.e.,
\begin{equation}
	y(kh) = x(kh)+\frac{C^*(q)}{D^*(q)}e(kh), \notag
\end{equation}
where $q$ denotes the forward shift operator, $e(kh)$ describes a zero-mean white noise stochastic process of finite variance that is uncorrelated with the input sequence $u(kh)$, and 
\begin{align}
	C^*(q)&= 1+c_1^*q^{-1}+c_2^*q^{-2}+\cdots + c_{m_c}^*q^{-m_c}, \notag \\
	D^*(q)&= 1+d_1^*q^{-1}+d_2^*q^{-2}+\cdots + d_{n_d}^*q^{-n_d}, \notag
\end{align}
with the polynomial degrees satisfying $n_d\geq m_c$. The coefficients $c_i^*$ and $d_j^*$ are combined in a parameter vector $\bm{\eta}^*$ of the same form as $\bm{\theta}_{\textnormal{c}}^*$ in \eqref{ctparametervector}.

Based on $N$ input and output data samples, $\{u(kh),y(kh)\}_{k=1}^N$, the goal is to determine a model for $G_\textnormal{c}^*(p):=B_\textnormal{c}^*(p)/A_\textnormal{c}^*(p)$ in the CT or DT domain, and possibly of the noise filter $H^*(q):=C^*(q)/D^*(q)$ also.


\subsection{Discrete-time equivalent and inverse of sampling}
Throughout this paper we assume that the input signal is constant between samples, i.e., it has a zero-order hold (ZOH) intersample behavior. Therefore, the system we intend to model can be exactly described at the sampling instants by its DT ZOH equivalent, which has the form
\begin{align}
	\label{dtsystem}
	y(kh)&= G_\textnormal{d}^*(q) u(kh) + H^*(q)e(kh),
\end{align}
with $G_\textnormal{d}^*(q)=B_\textnormal{d}^*(q)/A_\textnormal{d}^*(q)$, where
\begin{align}
	A_\textnormal{d}^*(q) &= \alpha_n^*q^n+\alpha_{n-1}^*q^{n-1}+\cdots + \alpha_1^*q + 1, \notag \\
	B_\textnormal{d}^*(q) &= \beta_n^*q^n+\beta_{n-1}^*q^{n-1}+\cdots + \beta_1^*q + \beta_0^*. \notag
\end{align}
The parameter vector associated with $G_\textnormal{d}^*(q)$ is denoted as $\bm{\theta}_\textnormal{d}^*$ and has the same structure as $\bm{\theta}_{\textnormal{c}}^*$ but is formed by the DT system parameters instead of the CT parameters. If the CT system is strictly proper, then $\beta_n^* = 0$ and the last element of $\bm{\theta}_{\textnormal{d}}^*$ is omitted. Note that, however, almost any strictly proper CT system leads to $\beta_{n-1}^*\neq 0$. Although this fact is well known \cite{yuz2014sampled}, explicit results on the resulting DT relative degree are difficult to find in the literature. For completeness, the formal statement with its proof is presented next.
\begin{proposition}
	\label{proposition1}
	Consider a strictly proper, LTI, CT system $G_\textnormal{c}^*(p)$. The relative degree of the DT ZOH equivalent of $G_\textnormal{c}^*(p)$ is $r\geq 1$ if and only if $y_\textnormal{c}^*(h)=y_\textnormal{c}^*(2h)=\cdots=y_\textnormal{c}^*([r-1]h)= 0$ and $y_\textnormal{c}^*(rh)\neq 0$, where $y_\textnormal{c}^*(t)$ is the step response of $G_\textnormal{c}^*(p)$ and $h$ is the sampling period.
\end{proposition}
\begin{proof}
	\hspace{-0.15cm} The DT ZOH equivalent of $\hspace{-0.04cm}G_\textnormal{c}^*(p)\hspace{-0.04cm}$ is computed~by
	\begin{align}
		G_{\textnormal{d}}^*(z) \hspace{-0.05cm} &= \hspace{-0.05cm}(1-z^{-1}) \mathcal{Z}\big\{y_\textnormal{c}^*(kh)\big\} \notag \\
		&= \hspace{-0.05cm}\frac{y_\textnormal{c}^*\hspace{-0.02cm}(h)}{z} \hspace{-0.05cm}+\hspace{-0.05cm} \frac{y_\textnormal{c}^*\hspace{-0.02cm}(2h)\hspace{-0.05cm}-\hspace{-0.05cm}y_\textnormal{c}^*\hspace{-0.02cm}(h)}{z^2}\hspace{-0.05cm}+\hspace{-0.05cm} \frac{y_\textnormal{c}^*\hspace{-0.02cm}(3h)\hspace{-0.05cm}-\hspace{-0.05cm}y_\textnormal{c}^*\hspace{-0.02cm}(2h)}{z^3}\hspace{-0.05cm}+\hspace{-0.05cm}\cdots\hspace{-0.02cm}. \notag 
	\end{align}
	Therefore, the transfer function $G_{\textnormal{d}}^*(z)$ has relative degree $r$ if and only if $y_\textnormal{c}^*(rh)-y_\textnormal{c}^*([r-1]h) \neq 0$ and
	\begin{equation}
		y_\textnormal{c}^*\hspace{-0.02cm}(h) \hspace{-0.05cm}=\hspace{-0.05cm} y_\textnormal{c}^*\hspace{-0.02cm}(2h)\hspace{-0.05cm}-\hspace{-0.05cm}y_\textnormal{c}^*(h) \hspace{-0.05cm}=\hspace{-0.06cm} \cdots \hspace{-0.06cm}=\hspace{-0.05cm} y_\textnormal{c}^*\hspace{-0.02cm}([r\hspace{-0.01cm}-\hspace{-0.01cm}1]h)\hspace{-0.05cm}-\hspace{-0.05cm}y_\textnormal{c}^*\hspace{-0.02cm}([r\hspace{-0.01cm}-\hspace{-0.01cm}2]h) \hspace{-0.05cm}=\hspace{-0.05cm} 0, \notag
	\end{equation}
	from which the statement follows.
\end{proof}

The indirect approach to CT system identification requires a link between the DT parameter vector $\bm{\theta}_\textnormal{d}$ and its CT equivalent $\bm{\theta}_{\textnormal{c}}$. This link is presented in Definition \ref{inversezoh}, which is used for analyzing the relationship between discrete and continuous-time refined instrumental variable methods.

\begin{definition}[Inverse ZOH transformation]
	\label{inversezoh}
	Given any DT transfer function described by $\bm{\theta}_{\textnormal{d}}$, we define~the \textit{inverse ZOH transformation} by $\mathbf{f}\colon \mathbb{R}^{2n+1}\to\mathbb{R}^{2n+1}; \bm{\theta}_{\textnormal{d}}\mapsto \bm{\theta}_{\textnormal{c}} = \mathbf{f}(\bm{\theta}_{\textnormal{d}})$. The vector $\mathbf{f}(\bm{\theta}_{\textnormal{d}})$ is of the form \eqref{ctparametervector} with $m=n$ that describes the transfer function $\mathbf{C}_\textnormal{c}(p\mathbf{I}-\mathbf{A}_\textnormal{c})^{-1}\mathbf{B}_\textnormal{c}+D_\textnormal{c}$, where 
	$\mathbf{C}_\textnormal{c}\hspace{-0.08cm} = \hspace{-0.1cm}\big[
	\beta_0\hspace{-0.03cm}-\hspace{-0.03cm}\tfrac{\beta_n}{\alpha_n}, \beta_1\hspace{-0.03cm}-\hspace{-0.03cm}\tfrac{\beta_n \alpha_1}{\alpha_n}, \hspace{-0.02cm}\dots\hspace{-0.02cm},\hspace{-0.02cm} \beta_{n\hspace{-0.02cm}-\hspace{-0.02cm}1}\hspace{-0.03cm}-\hspace{-0.03cm}\tfrac{\beta_n \alpha_{n-1}}{\alpha_n}\big], \hspace{0.07cm} D_\textnormal{c} \hspace{-0.08cm} = \hspace{-0.1cm} \beta_n/\hspace{-0.02cm}\alpha_n$, and where $\mathbf{A}_\textnormal{c}\in\mathbb{R}^{n\times n}$ with $\mathbf{B}_\textnormal{c}\in\mathbb{R}^{n\times 1}$ are computed by
	\vspace{-0.1cm}
	\begin{equation}
		\begin{bmatrix}
			\mathbf{A}_\textnormal{c} & \hspace{-0.2cm}\mathbf{B}_\textnormal{c} \\
			\mathbf{0} & \hspace{-0.2cm}0
		\end{bmatrix} \hspace{-0.1cm}=\hspace{-0.07cm} \frac{1}{h} \hspace{-0.02cm}\log\hspace{-0.03cm}\left(\begin{bmatrix}
			0 & \hspace{-0.1cm}1 & \hspace{-0.1cm}0 & \hspace{-0.1cm}\dots & \hspace{-0.1cm}0 & \hspace{-0.1cm}0 \\
			0 & \hspace{-0.1cm}0 & \hspace{-0.1cm}1 & \hspace{-0.1cm}\dots & \hspace{-0.1cm}0 & \hspace{-0.1cm}0 \\
			\vdots & \hspace{-0.1cm}\vdots & \hspace{-0.1cm}\vdots & \hspace{-0.1cm}\ddots & \hspace{-0.1cm}\vdots & \hspace{-0.1cm}\vdots \\
			0 & \hspace{-0.1cm}0 & \hspace{-0.1cm}0 & \hspace{-0.1cm}\dots & \hspace{-0.1cm}1 & \hspace{-0.1cm}0 \\
			\frac{-1}{\alpha_n} & \hspace{-0.1cm}\frac{-\alpha_1}{\alpha_n} & \hspace{-0.1cm}\frac{-\alpha_2}{\alpha_n} & \hspace{-0.1cm}\dots & \hspace{-0.1cm}\frac{-\alpha_{n-1}}{\alpha_n} & \hspace{-0.1cm}\frac{1}{\alpha_n} \\
			0 & \hspace{-0.1cm}0 & \hspace{-0.1cm}0 & \hspace{-0.1cm}\dots & \hspace{-0.1cm}0 & \hspace{-0.1cm}1
		\end{bmatrix} \right). \notag
	\end{equation}
In case $\beta_n=0$ (i.e., the DT model is strictly proper), we know that $b_n=0$ and we thus consider $\mathbf{f}$ to be the mapping between the strictly proper models only.
\end{definition}

The function $\mathbf{f}$ is known to be well-defined and bijective when 1) there is no negative real pole in the DT transfer function associated with $\bm{\theta}_{\textnormal{d}}$, and 2) the sampling radian frequency $\pi/h$ is larger than~twice the largest imaginary part of the poles related to $\mathbf{f}(\bm{\theta}_{\textnormal{d}})$ \cite{kollar1996equivalence}. The functions $\mathbf{f}$ and $\mathbf{f}^{-1}$ are well-defined and differentiable in the domain where these two conditions hold, thus leading to its Jacobian matrix $\partial \mathbf{f}/\partial \bm{\theta}_{\textnormal{d}}$ being non-singular at any point of that domain~\cite{rudin1976principles}.

\section{Refined instrumental variables: a unified notation}
\label{sec:RIV}
The motivation behind refined instrumental variables is to solve the non-convex optimization problem of maximum likelihood estimation by iterations stemming from a pseudo-linear regression. For brevity, similar to \cite{young2015refined}, here we jointly introduce the RIV (for DT identification) and RIVC (for CT identification) estimators via a unified notation.

We define $\xi$ as the unified operator (i.e., $q$ for the RIV estimator and $p$ for RIVC), and the unified subscript $(\cdot)_{\nu}$, where $\nu\in \{\textnormal{d},\textnormal{c}\}$ depending on whether the RIV or RIVC estimator is considered. Assuming that the $j$th iteration of the system parameters estimate, $\bm{\theta}_{\nu,j}$ (with model $B_{\nu,j}(\xi)/A_{\nu,j}(\xi)$), is given, the $j$th noise model iteration of the RIV and RIVC estimators is obtained by fitting an ARMA model to the estimated noise sequence $\{y(kh)-[B_{\nu,j}(\xi)/A_{\nu,j}(\xi)] u(kh)\}_{k=1}^N$. This procedure gives the noise model $C_{j+1}(q)/D_{j+1}(q)$, represented by the noise model parameter vector $\bm{\eta}_{j+1}$. 

There are several ways to fit an ARMA model to obtain $\bm{\eta}_{j+1}$. In this work we consider the prediction error method \cite{ljung1998system}. Formally, we define the function that links $\bm{\theta}_{\nu,j}$ and~$\bm{\eta}_{j\hspace{-0.02cm}+\hspace{-0.02cm}1}$ as $\mathbf{g}_\nu \colon \hspace{-0.04cm}\mathbb{R}^{n+m+1}\hspace{-0.05cm}\to \hspace{-0.05cm}\mathbb{R}^{m_c+n_d}; \bm{\theta}_{\nu,j}\mapsto \bm{\eta}_{j+1}=\mathbf{g}_{\nu}(\bm{\theta}_{\nu,j})$, where
\begin{equation}
	\mathbf{g}_{\nu}\hspace{-0.02cm}(\bm{\theta}_{\nu,j}\hspace{-0.01cm}) \hspace{-0.07cm}=\hspace{-0.04cm} \underset{\bm{\eta}}{\arg \min} \sum_{k=1}^N \hspace{-0.06cm} \left[\hspace{-0.02cm}\frac{D(q,\hspace{-0.02cm}\bm{\eta})}{C(q,\hspace{-0.02cm}\bm{\eta})}\hspace{-0.06cm}\left(\hspace{-0.06cm}y(kh)\hspace{-0.06cm}-\hspace{-0.07cm}\frac{B_{\nu,j}\hspace{-0.01cm}(\xi)}{A_{\nu,j}\hspace{-0.01cm}(\xi)}u(kh)\hspace{-0.04cm}\right)\hspace{-0.04cm}\right]^{\hspace{-0.03cm}2}\hspace{-0.1cm}. \notag
\end{equation}
\begin{remark}
	The notation $C(q,\bm{\eta})$ and $D(q,\bm{\eta})$ is used to stress the dependence of these polynomials on the parameter vector $\bm{\eta}$. Similarly, in the sequel we use $A_{\nu}(\xi,\bm{\theta}_{\nu,j})$ and $B_{\nu}(\xi,\bm{\theta}_{\nu,j})$ interchangeably for $A_{\nu,j}(\xi)$ and $B_{\nu,j}(\xi)$.
\end{remark}

After the noise model iteration is computed, an RIV step is proposed for minimizing $\mathbf{g}_{\nu}\hspace{-0.02cm}(\bm{\theta}_{\nu,j}\hspace{-0.01cm})$ upon convergence \cite{young2015refined}. The $j$th system model iteration is given by
\begin{align}
	\bm{\theta}_{\nu,j+1} &= \bigg[\frac{1}{N}\sum_{k=1}^{N}\hat{\bm{\varphi}}_{\nu,f}(kh,\bm{\theta}_{\nu,j}) \bm{\varphi}_{\nu,f}^\top(kh,\bm{\theta}_{\nu,j})  \bigg]^{-1} 	\notag \\
	\label{sriviterations}
	&\hspace{0.4cm}\times\bigg[\frac{1}{N}\sum_{k=1}^{N}\hat{\bm{\varphi}}_{\nu,f}(kh,\bm{\theta}_{\nu,j}) y_{\nu,f}(kh,\bm{\theta}_{\nu,j})  \bigg], 
\end{align}
where the matrix being inverted is called the \textit{modified normal matrix}. The filtered regressor $\bm{\varphi}_{\nu,f}(kh,\bm{\theta}_{\nu,j})$, filtered instrument $\hat{\bm{\varphi}}_{\nu,f}(kh,\bm{\theta}_{\nu,j})$ and filtered output $y_{\nu,f}(kh,\bm{\theta}_{\nu,j})$ are respectively\footnote{Note that $\bm{\varphi}_{\nu,f}, \hat{\bm{\varphi}}_{\nu,f}$ and $y_{\nu,f}$ have a subscript $f$ to stress the prefiltering process that must be performed. Also, note that for simplicity we have not written the dependence of these signals on the noise model.}given by

\begin{align}
	\bm{\varphi}_{\nu,f}(\hspace{-0.01cm}kh,\hspace{-0.025cm}\bm{\theta}_{\nu,j}\hspace{-0.02cm})\hspace{-0.08cm}&= \hspace{-0.09cm} \frac{D_{j\hspace{-0.01cm}+\hspace{-0.01cm}1}\hspace{-0.02cm}(q)}{C_{j\hspace{-0.01cm}+\hspace{-0.01cm}1}\hspace{-0.02cm}(q)}\hspace{-0.04cm}\bigg[
	\hspace{-0.05cm}\dfrac{-\xi}{A_{\nu,j}\hspace{-0.02cm}(\xi)} y(\hspace{-0.01cm}kh\hspace{-0.01cm}), \dots\hspace{-0.02cm}, \hspace{-0.06cm}\dfrac{-\xi^n}{A_{\nu,j}\hspace{-0.02cm}(\xi)} y(\hspace{-0.01cm}kh\hspace{-0.01cm}), \notag \\
	&\hspace{0.4cm}\dfrac{1}{A_{\nu,j}(\xi)}u(kh), \dots, \dfrac{\xi^m}{A_{\nu,j}(\xi)} u(kh) \bigg]^{\hspace{-0.04cm}\top}\hspace{-0.09cm}, \notag \\
	\hat{\bm{\varphi}}_{\nu,f}(kh,\hspace{-0.025cm}\bm{\theta}_{\nu,j}\hspace{-0.02cm}) \hspace{-0.07cm}&= \hspace{-0.09cm}\frac{D_{j\hspace{-0.01cm}+\hspace{-0.01cm}1}\hspace{-0.02cm}(q)}{C_{j\hspace{-0.01cm}+\hspace{-0.01cm}1}\hspace{-0.02cm}(q)}\hspace{-0.04cm}\bigg[ 
	\hspace{-0.03cm}\dfrac{-\xi B_{\nu,j}\hspace{-0.03cm}(\xi)}{A_{\nu,j}^2(\xi)} u(kh), \dots, \notag \\
	&\hspace{-1.3cm}\dfrac{-\xi^{n}\hspace{-0.04cm}B_{\nu,j}\hspace{-0.03cm}(\xi)}{A_{\nu,j}^2(\xi)} u(kh), \dfrac{1}{A_{\nu,j}(\xi)}u(kh), \dots, \dfrac{\xi^m}{A_{\nu,j}(\xi)} u(kh)
	\bigg]^{\hspace{-0.04cm}\top}\hspace{-0.09cm}, \notag \\
	\label{filteredoutput}
	y_{\nu,f}(kh,\hspace{-0.02cm}\bm{\theta}_{\nu,j}\hspace{-0.02cm}) \hspace{-0.07cm}&= \hspace{-0.09cm}  \frac{D_{j+1}(q)}{C_{j+1}(q)}\frac{1}{A_{\nu,j}(\xi)}y(kh).
\end{align}
In case the noise model is not estimated and thus set to unity (i.e., $C_j(q)=D_j(q)=1$) the iterations in \eqref{sriviterations} describe the SRIV and SRIVC estimators \cite{young1976some,young1980refined}.
\begin{remark}
	In the expressions above we have introduced a mixed notation of CT filters with sampled signals for the case $\nu=\textnormal{c}, \xi = p$ (i.e., for the RIVC and SRIVC estimators). If $G(p)$ is a CT filter and $x(kh)$ is a sampled signal, then $G(p)x(kh)$ implies that the signal $x(kh)$ is being interpolated using a ZOH, and the resulting output of the filter is sampled at $t=kh$. 
\end{remark}

Given the parameter estimate vector $\bar{\bm{\theta}}_{\textnormal{d}}$ obtained from the RIV iterations \eqref{sriviterations} at convergence as $j$ tends to infinity for a fixed $N$, the associated CT parameter vector (i.e., the indirect approach estimate) is simply given by $\mathbf{f}(\bm{\bar{\theta}}_{\textnormal{d}})$. 

\section{Relation between discrete-time and continuous-time refined instrumental variable methods}
\label{sec:equivalence}

Before stating the main theoretical results of this paper, we present two assumptions:
\begin{assumption}
	\label{assumption1}
	The system $G_{\textnormal{c}}^*(p)$ is proper ($n\geq m$) and asymptotically stable, with $A_{\textnormal{c}}^*(p)$ and $B_{\textnormal{c}}^*(p)$ being coprime.
\end{assumption}
\begin{assumption}
	\label{assumption2}
	For all $j$ sufficiently large, $A_{\textnormal{d},j}(q)$ has no negative real zero in the RIV algorithm, and the sampling frequency $\pi/h$ is larger than twice the largest imaginary part of the zeros of $A_{\textnormal{c},j}(p)$ in the RIVC algorithm.
\end{assumption}
While both assumptions are standard in the analysis of RIV algorithms (see, e.g., \cite{pan2020consistency,gonzalez2022theoretical}), the asymptotic stability of $G_{\textnormal{c}}^*(p)$ may not be needed in case the identification is performed in closed-loop, although a more involved filtering technique must be implemented \cite{gonzalez2022refined}. Furthermore, Assumption \ref{assumption2} is equivalent to requiring the inverse ZOH transformation $\mathbf{f}$ to be well defined along the iterative process. This will typically be the case if the sampling period is chosen adequately and the model iterations do not diverge.

Theorem \ref{thm1} shows that for models with relative degree zero or one, the RIV method with inverse ZOH transformation $\mathbf{f}$ provides the same model as the RIVC method upon convergence as the number of iterations $j$ tends to infinity for a fixed and finite sample size $N$.

\begin{theorem}
	\label{thm1}
	Consider the RIV and RIVC estimators described in Section \ref{sec:RIV}, and assume that upon convergence in iterations ($j\to \infty$) for a fixed sample size, their modified normal matrices are non-singular. Furthermore, assume that Assumptions \ref{assumption1} and \ref{assumption2} are satisfied, and that $m=n-1$ or $m=n$. Then, the RIVC and RIV methods have the same number of limiting points and they are linked by $\bar{\bm{\theta}}_{\textnormal{c}}=\mathbf{f}(\bar{\bm{\theta}}_{\textnormal{d}})$, where $\bar{\bm{\theta}}_{\textnormal{c}}$ and $\bar{\bm{\theta}}_{\textnormal{d}}$ are the limiting point(s) of the RIVC and RIV methods, respectively.
\end{theorem}
\begin{proof}
	Any limiting point $\bar{\bm{\theta}}_{\textnormal{d}}=\lim_{j\to \infty} \bm{\theta}_{\textnormal{d},j}$ of the RIV iterations must satisfy
	\begin{align}
		\bar{\bm{\theta}}_{\textnormal{d}} &= \bigg[\frac{1}{N}\sum_{k=1}^{N}\hat{\bm{\varphi}}_{\textnormal{d},f}(kh,\bar{\bm{\theta}}_{\textnormal{d}}) \bm{\varphi}_{\textnormal{d},f}^\top(kh,\bar{\bm{\theta}}_{\textnormal{d}})  \bigg]^{-1} 	\notag \\
		&\hspace{0.6cm}\times\bigg[\frac{1}{N}\sum_{k=1}^{N}\hat{\bm{\varphi}}_{\textnormal{d},f}(kh,\bar{\bm{\theta}}_{\textnormal{d}}) y_{\textnormal{d},f}(kh,\bar{\bm{\theta}}_{\textnormal{d}}) \bigg]. \notag 
	\end{align}
	Provided the modified normal matrix is non-singular, the condition above reduces to
	\begin{equation}
		\label{reducesto}
		\sum_{k=1}^{N}\hat{\bm{\varphi}}_{\textnormal{d},f}(kh,\bar{\bm{\theta}}_{\textnormal{d}}) \left[y_{\textnormal{d},f}(kh,\bar{\bm{\theta}}_{\textnormal{d}})-\bm{\varphi}_{\textnormal{d},f}^\top(kh,\bar{\bm{\theta}}_{\textnormal{d}})\bar{\bm{\theta}}_{\textnormal{d}} \right] =\mathbf{0}.
	\end{equation}
	After some algebraic manipulations, we can see that
	\begin{equation}
		\label{expression1}
	\hat{\bm{\varphi}}_{\textnormal{d},\hspace{-0.02cm}f}\hspace{-0.03cm}(kh,\hspace{-0.02cm}\bar{\bm{\theta}}_{\textnormal{d}})\hspace{-0.05cm}=\hspace{-0.05cm} \frac{D(q,\mathbf{g}_{\textnormal{d}}(\bar{\bm{\theta}}_{\textnormal{d}}))}{C(q,\mathbf{g}_{\textnormal{d}}(\bar{\bm{\theta}}_{\textnormal{d}}))} \frac{\partial }{\partial \bm{\theta}_{\textnormal{d}}} \frac{B_{\textnormal{d}}(q,\bm{\theta}_{\textnormal{d}})}{A_{\textnormal{d}}(q,\bm{\theta}_{\textnormal{d}})}u(kh)\bigg|_{\bm{\theta}_{\textnormal{d}}=\bar{\bm{\theta}}_{\textnormal{d}}}\hspace{-0.07cm},\hspace{-0.05cm}
	\end{equation}
	and
	\begin{align}
		y_{\textnormal{d},f}(kh,&\bar{\bm{\theta}}_{\textnormal{d}})-\bm{\varphi}_{\textnormal{d},f}^\top(kh,\bar{\bm{\theta}}_{\textnormal{d}})\bar{\bm{\theta}}_{\textnormal{d}} = \notag \\
		\label{expression2}
		&\frac{D(q,\mathbf{g}_{\textnormal{d}}(\bar{\bm{\theta}}_{\textnormal{d}}))}{C(q,\mathbf{g}_{\textnormal{d}}(\bar{\bm{\theta}}_{\textnormal{d}}))}\left(y(kh)-\frac{B_{\textnormal{d}}(q,\bar{\bm{\theta}}_{\textnormal{d}})}{A_{\textnormal{d}}(q,\bar{\bm{\theta}}_{\textnormal{d}})}u(kh)\right).
	\end{align}
	Replacing \eqref{expression1} and \eqref{expression2} in \eqref{reducesto} leads to 
	\begin{align}
		\sum_{k=1}^{N}&\frac{D(q,\mathbf{g}_{\textnormal{d}}(\bar{\bm{\theta}}_{\textnormal{d}}))}{C(q,\mathbf{g}_{\textnormal{d}}(\bar{\bm{\theta}}_{\textnormal{d}}))} \frac{\partial }{\partial \bm{\theta}_{\textnormal{d}}} \frac{B_{\textnormal{d}}(q,\bm{\theta}_{\textnormal{d}})}{A_{\textnormal{d}}(q,\bm{\theta}_{\textnormal{d}})}u(kh)\bigg|_{\bm{\theta}_{\textnormal{d}}=\bar{\bm{\theta}}_{\textnormal{d}}} \times \notag \\
		\label{conditionsriv}
		&\hspace{0.2cm}\frac{D(q,\mathbf{g}_{\textnormal{d}}(\bar{\bm{\theta}}_{\textnormal{d}}))}{C(q,\mathbf{g}_{\textnormal{d}}(\bar{\bm{\theta}}_{\textnormal{d}}))}\left(y(kh)-\frac{B_{\textnormal{d}}(q,\bar{\bm{\theta}}_{\textnormal{d}})}{A_{\textnormal{d}}(q,\bar{\bm{\theta}}_{\textnormal{d}})}u(kh)\right) = \mathbf{0}.
	\end{align}
	The existence of the inverse ZOH transformation for $\bar{\bm{\theta}}_\textnormal{d}$ is ensured by Assumption \ref{assumption2}. Since the ZOH equivalence relation is exact at the sampling instants \cite{astroem1984computer}, the equality $[B_{\textnormal{d}}(q,\bar{\bm{\theta}}_{\textnormal{d}})/A_{\textnormal{d}}(q,\bar{\bm{\theta}}_{\textnormal{d}})]u(kh) = [B_{\textnormal{c}}(p,\mathbf{f}(\bar{\bm{\theta}}_{\textnormal{d}}))/A_{\textnormal{c}}(p,\mathbf{f}(\bar{\bm{\theta}}_{\textnormal{d}}))]u(kh)$ holds for all $k\in \mathbb{N}$. By the same argument, we find that $\mathbf{g}_{\textnormal{d}}(\bar{\bm{\theta}}_\textnormal{d})=\mathbf{g}_{\textnormal{c}}(\mathbf{f}(\bar{\bm{\theta}}_\textnormal{d}))$. Moreover, the chain rule for gradients yields
	\begin{align}
		\frac{\partial }{\partial \bm{\theta}_{\textnormal{d}}} \frac{B_{\textnormal{d}}(q,\bm{\theta}_{\textnormal{d}})}{A_{\textnormal{d}}(q,\bm{\theta}_{\textnormal{d}})}u(kh)\bigg|_{\bm{\theta}_{\textnormal{d}}=\bar{\bm{\theta}}_{\textnormal{d}}} &= \frac{\partial }{\partial \bm{\theta}_{\textnormal{d}}} \frac{B_{\textnormal{c}}(p,\mathbf{f}(\bm{\theta}_{\textnormal{d}}))}{A_{\textnormal{c}}(p,\mathbf{f}(\bm{\theta}_{\textnormal{d}}))}u(kh)\bigg|_{\bm{\theta}_{\textnormal{d}}=\bar{\bm{\theta}}_{\textnormal{d}}} \notag \\
		\label{gradient}
		&\hspace{-2.5cm}=\frac{\partial \mathbf{f}}{\partial \bm{\theta}_{\textnormal{d}}}\bigg|_{\bm{\theta}_{\textnormal{d}}=\bar{\bm{\theta}}_{\textnormal{d}}} \frac{\partial }{\partial \bm{\theta}_{\textnormal{c}}} \frac{B_{\textnormal{c}}(p,\bm{\theta}_{\textnormal{c}})}{A_{\textnormal{c}}(p,\bm{\theta}_{\textnormal{c}})}u(kh)\bigg|_{\bm{\theta}_{\textnormal{c}}=\mathbf{f}(\bar{\bm{\theta}}_{\textnormal{d}})}\hspace{-0.05cm}. 
	\end{align}
	Note that $\partial \mathbf{f}/\partial \bm{\theta}_{\textnormal{d}}$ is non-singular thanks to $\mathbf{f}$ and $\mathbf{f}^{-1}$ being continuous and differentiable in the domain of interest due to Assumption \ref{assumption2}. Such matrix does not depend on $N$ and can therefore be factored out of the sum in \eqref{conditionsriv} thanks to the linearity of the transfer functions. Thus, \eqref{conditionsriv} is equivalent to
	\begin{align}
		\sum_{k=1}^{N}&\frac{D(q,\mathbf{g}_{\textnormal{c}}(\mathbf{f}(\bar{\bm{\theta}}_\textnormal{d})))}{C(q,\mathbf{g}_{\textnormal{c}}(\mathbf{f}(\bar{\bm{\theta}}_\textnormal{d})))} \frac{\partial }{\partial \bm{\theta}_{\textnormal{c}}} \frac{B_{\textnormal{c}}(p,\bm{\theta}_{\textnormal{c}})}{A_{\textnormal{c}}(p,\bm{\theta}_{\textnormal{c}})}u(kh)\bigg|_{\bm{\theta}_{\textnormal{c}}=\mathbf{f}(\bar{\bm{\theta}}_{\textnormal{d}})} \times \notag \\
		&\frac{D(q,\mathbf{g}_{\textnormal{c}}(\mathbf{f}(\bar{\bm{\theta}}_\textnormal{d})))}{C(q,\mathbf{g}_{\textnormal{c}}(\mathbf{f}(\bar{\bm{\theta}}_\textnormal{d})))}\left(y(kh)-\frac{B_{\textnormal{c}}(p,\mathbf{f}(\bar{\bm{\theta}}_{\textnormal{d}}))}{A_{\textnormal{c}}(p,\mathbf{f}(\bar{\bm{\theta}}_{\textnormal{d}}))}u(kh)\right) = \mathbf{0}. \notag
	\end{align}
	On the other hand, by following the same arguments to derive \eqref{conditionsriv} but for the RIVC method, any limiting point $\bar{\bm{\theta}}_{\textnormal{c}}$ satisfies
	\begin{align}
	\sum_{k=1}^{N}&\frac{D(q,\mathbf{g}_{\textnormal{c}}(\bar{\bm{\theta}}_{\textnormal{c}}))}{C(q,\mathbf{g}_{\textnormal{c}}(\bar{\bm{\theta}}_{\textnormal{c}}))} \frac{\partial }{\partial \bm{\theta}_{\textnormal{c}}} \frac{B_{\textnormal{c}}(p,\bm{\theta}_{\textnormal{c}})}{A_{\textnormal{c}}(p,\bm{\theta}_{\textnormal{c}})}u(kh)\bigg|_{\bm{\theta}_{\textnormal{c}}=\bar{\bm{\theta}}_{\textnormal{c}}} \times \notag \\
	&\frac{D(q,\mathbf{g}_{\textnormal{c}}(\bar{\bm{\theta}}_{\textnormal{c}}))}{C(q,\mathbf{g}_{\textnormal{c}}(\bar{\bm{\theta}}_{\textnormal{c}}))}\left(y(kh)-\frac{B_{\textnormal{c}}(p,\bar{\bm{\theta}}_{\textnormal{c}})}{A_{\textnormal{c}}(p,\bar{\bm{\theta}}_{\textnormal{c}})}u(kh)\right) = \mathbf{0}. \notag
	\end{align}
	By comparing the characterizations of $\bar{\bm{\theta}}_\textnormal{d}$ and $\bar{\bm{\theta}}_\textnormal{c}$, we conclude that the limiting points of the RIV and RIVC estimators are linked by $\bar{\bm{\theta}}_{\textnormal{c}}=\mathbf{f}(\bar{\bm{\theta}}_{\textnormal{d}})$.
\end{proof}
\begin{corollary}
	Under the same assumptions as in Theorem \ref{thm1}, the SRIVC and SRIV methods have the same number of limiting points and they are linked by $\bar{\bm{\theta}}_{\textnormal{c}}=\mathbf{f}(\bar{\bm{\theta}}_{\textnormal{d}})$, where $\bar{\bm{\theta}}_{\textnormal{c}}$ and $\bar{\bm{\theta}}_{\textnormal{d}}$ are the limiting point(s) of the SRIVC and SRIV methods, respectively.
\end{corollary}
\begin{proof}
	Direct by fixing $C_j(q)=D_j(q)=1$ in the proof for Theorem \ref{thm1}.
\end{proof}

\begin{remark}
	The non-singularity of the modified normal matrices of the refined instrumental variable methods depends on the persistence of excitation order of the input, as well as the amount of over-parametrization, if any \cite{gonzalez2022theoretical}. General conditions for the non-singularity of this matrix have not been addressed in the literature, although sufficient conditions for the generic non-singularity of such matrix for the SRIVC method can be found in \cite{pan2020consistency}. Similar conditions can be derived for the SRIV estimator, although they are outside the scope of this work.
\end{remark}

Theorem \ref{thm1} shows that, under similar initialization conditions, the indirect and direct approaches with refined instrumental variables approach the \textit{same} estimate at convergence in iterations when $m=n-1$ or $m=n$. This theorem only applies for such values of $m$ since only in these two cases there is a bijection between the CT and DT models (corresponding to $\textbf{f}$ or its reduced version). In such scenarios, one could argue that it is useless to perform the RIVC iterations with the RIV estimate as an initialization point, as it is done in some applications of the RIVC method \cite{ljung2009experiments}. However, it has been noted in \cite{garnier2014advantages} that the CT equivalent of the RIV estimator can fail to deliver reliable models, usually in the presence of fast sampling or stiff systems. Possible discrepancies between both methods when $m=n-1$ or $m=n$ can be explained by
\begin{itemize}
	\item Ill-conditioning of the Jacobian $\partial \mathbf{f}/\partial \bm{\theta}_{\textnormal{d}}$: for stiff systems this matrix can be severely ill-conditioned, which can affect the convergence of the iterative procedure of the RIV estimator. Also, instability of the estimates can arise more frequently within the iterations, thus requiring ad-hoc stabilization steps \cite{gonzalez2022refined}.
	\item Misspecification of the intersample behavior in the RIVC method: If the intersample behavior used for prefiltering the input in the RIVC method does not match with the nature of the DT equivalents in the RIV estimator, then both methods will deliver different results in general.
	\item Choice of initialization: Convergence is not guaranteed for finite samples, and the estimators may converge to (different) local minima if poorly initialized. 
\end{itemize}
\subsection{The Adapted RIV estimator}
The indirect and direct approaches no longer produce the same estimates for $m<n-1$, since they propose different model structures. It has been noted in \cite{ljung2009experiments} that the indirect approach will typically lead to worse results when $m<n-1$. However, it is possible to establish a link between the indirect and direct procedures if the filtered instrument and regressor vectors of the RIV method are appropriately modified. In the sequel, we consider the system parametrization
\begin{equation}
	G_{\textnormal{d}}^*(q) = \frac{\sum_{i=0}^m N_{\textnormal{d},i}^*(q) \gamma_i^*}{A_{\textnormal{d}}^*(q)} \notag 
\end{equation}
with parameter vector $\bm{\rho}_{\textnormal{d}}^*= [\alpha_1^*, \dots, \alpha_n^*, \gamma_0^*, \dots, \gamma_m^*]^\top$, and where $N_{\textnormal{d},i}^*(q)$ are the numerator polynomials of the DT equivalents of $p^i/A_{\textnormal{c}}^*(p)$. This novel parametrization is such that its numerator coefficients $\gamma_i^*$ correspond exactly to those of the numerator polynomial of its CT equivalent. By leveraging the state-space description of $p^i/A_{\textnormal{c}}^*(p)$, it is possible to show that 
\begin{equation}
	N_{\textnormal{d},i}^*(q) = \alpha_n^* \mathbf{e}_{i+1}^\top \textnormal{adj}(q\mathbf{I}-e^{\mathbf{A}_{\textnormal{c}}^*h})(\mathbf{I}-e^{\mathbf{A}_{\textnormal{c}}^*h})\mathbf{e}_1, \notag
\end{equation}
where $\textnormal{adj}(\cdot)$ denotes the adjugate matrix, $\mathbf{e}_j$ is the $j$th column of the identity matrix of appropriate size, and $\mathbf{A}_\textnormal{c}^*$ is the state matrix of the CT equivalent of $G_{\textnormal{d}}^*(q)$, written as
\begin{equation}
	\mathbf{A}_{\textnormal{c}}^* = \begin{bmatrix}
		0 & \hspace{-0.1cm}1 & \hspace{-0.1cm}0 & \hspace{-0.1cm}\dots & \hspace{-0.1cm}0  \\
		0 & \hspace{-0.1cm}0 & \hspace{-0.1cm}1 & \hspace{-0.1cm}\dots & \hspace{-0.1cm}0  \\
		\vdots & \hspace{-0.1cm}\vdots & \hspace{-0.1cm}\vdots & \hspace{-0.1cm}\ddots & \hspace{-0.1cm}\vdots  \\
		0 & \hspace{-0.1cm}0 & \hspace{-0.1cm}0 & \hspace{-0.1cm}\dots & \hspace{-0.1cm}1  \\
		\frac{-1}{a_n^*} & \hspace{-0.1cm}\frac{-a_1^*}{a_n^*} & \hspace{-0.1cm}\frac{-a_2^*}{a_n^*} & \hspace{-0.1cm}\dots & \hspace{-0.1cm}\frac{-a_{n-1}^*}{a_n^*} 
	\end{bmatrix}. \notag
\end{equation}
Note that the $N_{\textnormal{d},i}^*(q)$ polynomials are in general of order $n\hspace{-0.018cm}-\hspace{-0.018cm}1$ due to Proposition \ref{proposition1}, and they satisfy $N_{\textnormal{d},i}^*(1)\hspace{-0.04cm}=\hspace{-0.04cm}0$ for $i>0$. The vector $\bm{\rho}_{\textnormal{d}}^*$ is related to $\bm{\theta}_{\textnormal{c}}^*$ via $\mathbf{f}$ for the~de\-nominator coefficients, and via the identity map for the numerator coefficients. This transformation will be denoted as $\tilde{\mathbf{f}}$ and it is bijective in the same domain as $\mathbf{f}$ in Definition \ref{inversezoh}.

\begin{definition}[Adapted RIV and SRIV estimators]
	\label{defadaptedsriv}
	We define the Adapted RIV (ARIV) estimator with noise model iterations given by $\bm{\eta}_{j+1}=\tilde{\mathbf{g}}_{\textnormal{d}}(\bm{\rho}_{\textnormal{d},j})$, with
	\begin{align}
	&\tilde{\mathbf{g}}_{\textnormal{d}}(\bm{\rho}_{\textnormal{d},j}) = \underset{\bm{\eta}}{\arg \min} \sum_{k=1}^N  \bigg[\frac{D(q,\bm{\eta})}{C(q,\bm{\eta})}\notag \\
	\label{gtilde}
	&\hspace{0.5cm}\times\left(y(kh)-\frac{\sum_{i=0}^m N_{\textnormal{d},i}(q,\bm{\rho}_{\textnormal{d},j}) \gamma_{i,j}}{A_{\textnormal{d}}(q,\bm{\rho}_{\textnormal{d},j})}u(kh)\right)\bigg]^{2},
	\end{align}
	and system model iterations \eqref{sriviterations} with filtered output \eqref{filteredoutput}, but where the filtered regressor and instrument vectors are, respectively, given by
	\begin{align}
		\hspace{-0.07cm}\bm{\varphi}_{\textnormal{d},f}(kh,\bm{\rho}_{\textnormal{d},j}\hspace{-0.02cm})\hspace{-0.07cm}&= \hspace{-0.09cm} \frac{D_{j\hspace{-0.01cm}+\hspace{-0.01cm}1}\hspace{-0.02cm}(q)}{C_{j\hspace{-0.01cm}+\hspace{-0.01cm}1}\hspace{-0.02cm}(q)}\hspace{-0.04cm}\bigg[
		\hspace{-0.04cm}\dfrac{-q}{A_{\textnormal{d},j}\hspace{-0.02cm}(q)} y(\hspace{-0.01cm}kh\hspace{-0.01cm}), \dots, \hspace{-0.04cm}\dfrac{-q^n}{A_{\textnormal{d},j}\hspace{-0.02cm}(q)} y(\hspace{-0.01cm}kh\hspace{-0.01cm}), \notag \\
		\label{regressorsrivadapted}
		&\hspace{-0.35cm}\dfrac{N_{\textnormal{d},0,j}(q)}{A_{\textnormal{d},j}(q)}u(kh), \dots, \dfrac{N_{\textnormal{d},m,j}(q)}{A_{\textnormal{d},j}(q)} u(kh) \bigg]^{\hspace{-0.04cm}\top}\hspace{-0.04cm},\\
		\hat{\bm{\varphi}}_{\textnormal{d},\hspace{-0.02cm}f}(kh,\hspace{-0.02cm}\bm{\rho}_{\textnormal{d},\hspace{-0.02cm}j}\hspace{-0.02cm}) \hspace{-0.08cm}&= \hspace{-0.07cm}\frac{D_{j+1}(q)}{C_{j+1}(q)} \bigg[ 
		\dfrac{-M_{\textnormal{d},1,j}(q)}{A_{\textnormal{d},j}^2(q)} u(kh), \dots, \notag \\
		\label{instrumentsrivadapted}
		&\hspace{-1.8cm}\dfrac{-\hspace{-0.04cm}M_{\hspace{-0.02cm}\textnormal{d},\hspace{-0.02cm}n\hspace{-0.01cm},\hspace{-0.01cm}j}\hspace{-0.03cm}(\hspace{-0.01cm}q\hspace{-0.01cm})}{A_{\textnormal{d},j}^2(q)} u(\hspace{-0.01cm}kh\hspace{-0.01cm}),\dfrac{N_{\textnormal{d},0,j}(q)}{A_{\textnormal{d},j}(q)}u(kh), \dots, \dfrac{N_{\textnormal{d},m,j}(q)}{A_{\textnormal{d},j}(q)} u(kh) \bigg]^{\hspace{-0.04cm}\top}\hspace{-0.1cm}, 
	\end{align}
	with $N_{\textnormal{d},l,j}(q)$ and $M_{\textnormal{d},r,j}(q)$ ($l=0,\dots,m$, $r=1,\dots,n$) being, respectively, the numerator polynomials of the DT equivalents of $p^l/A_{\textnormal{c}}(p,\tilde{\mathbf{f}}(\bm{\rho}_{\textnormal{d},j}))$ and
	\begin{equation}
		[p, \dots, p^n, 0,\dots,0] \frac{\partial \tilde{\mathbf{f}}(\bm{\rho}_{\textnormal{d}})}{\partial\alpha_r}\bigg|_{\bm{\rho}_{\textnormal{d}}=\bm{\rho}_{\textnormal{d},j}} \frac{B_{\textnormal{c}}(p,\tilde{\mathbf{f}}(\bm{\rho}_{\textnormal{d},j}))}{A_{\textnormal{c}}^2 (p,\tilde{\mathbf{f}}(\bm{\rho}_{\textnormal{d},j}))}. \notag 
	\end{equation}
	Upon convergence in iterations, the resulting DT model is given by $(\sum_{i=0}^m N_{\textnormal{d},i}(q,\bar{\bm{\rho}}_{\textnormal{d}}) \bar{\gamma}_i)/A_{\textnormal{d}}(q,\bar{\bm{\rho}}_{\textnormal{d}})$, where $\bar{\bm{\rho}}_{\textnormal{d}}=\lim_{j\to\infty} \bm{\rho}_{\textnormal{d},j}$ and $\bar{\gamma}_i=\lim_{j\to\infty}\gamma_{i,j}$. The Adapted SRIV (ASRIV) estimator is defined as the ARIV estimator but with a fixed noise model $C_j(q)=D_j(q)=1$.
\end{definition}

\begin{theorem}
	\label{thm2}
	Consider the ARIV algorithm in Definition \ref{defadaptedsriv} and the RIVC estimator in Section \ref{sec:RIV}, and assume that upon convergence in iterations for a fixed sample size, their modified normal matrices are non-singular. If Assumptions \ref{assumption1} and \ref{assumption2} are satisfied, then the ARIV and RIVC estimators have the same number of limiting points and they are linked by  $\bar{\bm{\theta}}_{\textnormal{c}}=\tilde{\mathbf{f}}(\bar{\bm{\rho}}_{\textnormal{d}})$, where $\bar{\bm{\theta}}_{\textnormal{c}}$ and $\bar{\bm{\rho}}_{\textnormal{d}}$ are the limiting point(s) of the RIVC and ARIV methods, respectively.
\end{theorem}
\begin{proof}
	At convergence, a limiting point $\bar{\bm{\rho}}_{\textnormal{d}}$ of the ARIV iterations must satisfy \eqref{reducesto}, however now
	\begin{align}
		y_{\textnormal{d},f}&(kh,\bar{\bm{\rho}}_{\textnormal{d}})-\bm{\varphi}_{\textnormal{d},f}^\top(kh,\bar{\bm{\rho}}_{\textnormal{d}})\bar{\bm{\rho}}_{\textnormal{d}}\notag \\
		&=\frac{D(q,\tilde{\mathbf{g}}_{\textnormal{d}}(\bar{\bm{\rho}}_{\textnormal{d}}))}{C(q,\tilde{\mathbf{g}}_{\textnormal{d}}(\bar{\bm{\rho}}_{\textnormal{d}}))}\left(y(kh)-\frac{\sum_{i=0}^m N_{\textnormal{d},i}(q,\bar{\bm{\rho}}_{\textnormal{d}}) \bar{\gamma}_i}{A_{\textnormal{d}}(q,\bar{\bm{\rho}}_{\textnormal{d}})}u(kh)\right) \notag \\
		&=\frac{D(q,\mathbf{g}_{\textnormal{c}}(\tilde{\mathbf{f}}(\bar{\bm{\rho}}_{\textnormal{d}})))}{C(q,\mathbf{g}_{\textnormal{c}}(\tilde{\mathbf{f}}(\bar{\bm{\rho}}_{\textnormal{d}})))}\left(y(kh)-\frac{B_{\textnormal{c}}(p,\tilde{\mathbf{f}}(\bar{\bm{\rho}}_{\textnormal{d}}))}{A_{\textnormal{c}}(p,\tilde{\mathbf{f}}(\bar{\bm{\rho}}_{\textnormal{d}}))}u(kh)\right). \notag
	\end{align}
	After some algebraic manipulations, it can be noted that the entries of $\hat{\bm{\varphi}}_{\textnormal{d},f}(kh,\bar{\bm{\rho}}_{\textnormal{d}})$ satisfy
	\begin{equation}
		\dfrac{-M_{\textnormal{d},r}(q,\bar{\bm{\rho}}_{\textnormal{d}})}{A_{\textnormal{d}}^2(q,\bar{\bm{\rho}}_{\textnormal{d}})} u(kh) = \frac{\partial }{\partial \alpha_r} \frac{B_{\textnormal{c}}(p,\tilde{\mathbf{f}}(\bm{\rho}_{\textnormal{d}}))}{A_{\textnormal{c}}(p,\tilde{\mathbf{f}}(\bm{\rho}_{\textnormal{d}}))} \bigg|_{\bm{\rho}_{\textnormal{d}} = \bar{\bm{\rho}}_{\textnormal{d}}} \notag
	\end{equation}
	for $r=1,\dots,n$, and
	\begin{align}
		\dfrac{N_{\textnormal{d},l}(q,\bar{\bm{\rho}}_{\textnormal{d}})}{A_{\textnormal{d}}(q,\bar{\bm{\rho}}_{\textnormal{d}})}u(kh) &= \frac{\partial }{\partial \gamma_l} \frac{\sum_{i=0}^m  N_{\textnormal{d},i}(q,\bm{\rho}_{\textnormal{d}}) \gamma_i}{A_{\textnormal{d}}(q,\bm{\rho}_{\textnormal{d}})}u(kh) \bigg|_{\bm{\rho}_{\textnormal{d}}= \bar{\bm{\rho}}_{\textnormal{d}}} \notag \\
		&= \frac{\partial }{\partial \gamma_l} \frac{B_{\textnormal{c}}(p,\tilde{\mathbf{f}}(\bm{\rho}_{\textnormal{d}}))}{A_{\textnormal{c}}(p,\tilde{\mathbf{f}}(\bm{\rho}_{\textnormal{d}}))} \bigg|_{\bm{\rho}_{\textnormal{d}} = \bar{\bm{\rho}}_{\textnormal{d}}} \notag
	\end{align}
	for $l=1,\dots,m$. Therefore, by the chain rule for gradients (cf. \eqref{gradient}),
	\begin{align}
		\hat{\bm{\varphi}}_{\textnormal{d},f}&(kh,\bar{\bm{\rho}}_{\textnormal{d}}) = \notag \\
		&\hspace{-0.1cm}\frac{D(q,\mathbf{g}_{\textnormal{c}}(\tilde{\mathbf{f}}(\bar{\bm{\rho}}_{\textnormal{d}})))}{C(q,\mathbf{g}_{\textnormal{c}}(\tilde{\mathbf{f}}(\bar{\bm{\rho}}_{\textnormal{d}})))}\frac{\partial \tilde{\mathbf{f}}}{\partial \bm{\rho}_{\textnormal{d}}}\bigg|_{\bm{\rho}_{\textnormal{d}}=\bar{\bm{\rho}}_{\textnormal{d}}}\frac{\partial }{\partial \bm{\theta}_{\textnormal{c}}} \frac{B_{\textnormal{c}}(p,\bm{\theta}_{\textnormal{c}})}{A_{\textnormal{c}}(p,\bm{\theta}_{\textnormal{c}})}u(kh)\bigg|_{\bm{\theta}_{\textnormal{c}}=\tilde{\mathbf{f}}(\bar{\bm{\rho}}_{\textnormal{d}})}\hspace{-0.07cm}. \notag
	\end{align}
	The rest of the proof follows the same lines as the proof of Theorem \ref{thm1} after \eqref{gradient} and is therefore omitted.
\end{proof}	
\begin{corollary}
	Under the same assumptions as in Theorem \ref{thm2}, the ASRIV and the SRIVC estimators have the same number of limiting points, and they are linked by $\bar{\bm{\theta}}_{\textnormal{c}}=\tilde{\mathbf{f}}(\bar{\bm{\rho}}_{\textnormal{d}})$, where $\bar{\bm{\theta}}_{\textnormal{c}}$ and $\bar{\bm{\rho}}_{\textnormal{d}}$ are the limiting point(s) of the SRIVC and ASRIV methods, respectively.
\end{corollary}
\begin{proof}
	Direct from fixing $C_j(q)=D_j(q)=1$ in the proof for Theorem \ref{thm2}.
\end{proof}

In summary, the ARIV method in Definition \ref{defadaptedsriv} provides a DT estimator whose CT equivalent model has a fixed relative degree $n-m$. This implementation has the advantage of enforcing smoothness properties of the CT system directly into the DT estimate, without the need of additional optimization steps such as in \cite{gonzalez2018asymptotically}. The proposed method closes the gap between DT and CT refined instrumental variable methods, since the CT equivalent of the ARIV estimator corresponds exactly to the RIVC estimator for a finite sample size.

\begin{table*}[t]
	\caption{Mean square errors of the parameters for the methods SRIV and RIV, and their adapted counterparts ASRIV and ARIV.}
	\centering
	\begin{tabular}{c|c|c|c|c|c|c|c|c|c|c}
		Method & $\alpha_1$ & $\alpha_2$ & $\alpha_3$ & $\alpha_4$ & $\beta_0$ & $\beta_1$ & $\beta_2$ & $\beta_3$ & $d_1$   & $c_1$     \\ \hline
		SRIV   & 4.06e-5 & 3.05e-4 & 3.16e-4 & 4.81e-5 & 9.61e-4 & 4.25e-3 & 3.40e-3 & 5.59e-4 & - & - \\
		ASRIV  & 2.46e-5 & 2.50e-4 & 3.64e-4 & 6.71e-5 & 9.69e-6 & 1.05e-4 & 8.79e-5 & 1.56e-5 & - & - \\
		RIV    & 3.97e-5 & 3.00e-4 & 3.15e-4 & 4.81e-5 & 9.58e-4 & 4.25e-3 & 3.42e-3 & 5.60e-4 & 6.83e-5 & 1.13e-4 \\ 
		ARIV   & 2.23e-5 & 2.02e-4 & 2.65e-4 & 4.55e-5 & 4.84e-6 & 5.32e-5 & 4.39e-5 & 7.91e-6 & 6.82e-5 & 1.13e-4
	\end{tabular}
\vspace{-0.5cm}
	\label{table1}
\end{table*}

\subsection{The Adapted RIVC estimator}

The analog problem of the previous subsection for CT system identification consists of directly identifying a CT system whose DT equivalent has relative degree $n-m>1$. This problem arises if the system is known to have a fixed number of sample time-delays. Instead of computing a time-delayed CT system, which typically requires iterative non-convex optimization steps
, a computationally-cheap approach that directly imposes sample time-delays in the estimated transfer function can be obtained by an adapted form of the RIVC estimator, similar to Definition \ref{defadaptedsriv}. In this case, we must parameterize the CT system as 
\begin{equation}
G_{\textnormal{c}}^*(p) = \frac{\sum_{i=0}^m N_{\textnormal{c},i}^*(p) \nu_i^*}{A_{\textnormal{c}}^*(p)} \notag
\end{equation}
with parameter vector $\bm{\rho}_{\textnormal{c}}^*= [a_1^*, \dots, a_n^*, \nu_0^*, \dots, \nu_m^*]^\top$, and where $N_{\textnormal{c},i}^*(p)$ are the numerator polynomials of the CT equivalent of $q^i/A_{\textnormal{d}}^*(q)$. The noise model is computed from an ARMA estimation step similar to \eqref{gtilde}, while the system parameter vector $\bm{\rho}_{\textnormal{c}}^*$ can be estimated from a refined instrumental variable procedure with filtered regressor and instrument vectors given by CT variants of \eqref{regressorsrivadapted} and \eqref{instrumentsrivadapted}. We omit the full expressions due to space~constraints.

\section{Simulation example}
\label{simulations}
To provide an example of the theoretical results, we consider the Rao-Garnier benchmark system \cite{rao2002numerical}
\begin{equation}
	G_\textnormal{c}^*(p) = \frac{-4p+1}{0.000625p^4\hspace{-0.02cm}+0.003125p^3\hspace{-0.02cm}+0.255p^2\hspace{-0.02cm}+0.26p+1}, \notag
\end{equation}
with noise model $H^*\hspace{-0.03cm}(q)\hspace{-0.04cm} =\hspace{-0.04cm} (1\hspace{-0.02cm}+\hspace{-0.02cm}0.4q^{\hspace{-0.02cm}-\hspace{-0.02cm}1}\hspace{-0.02cm})/(1\hspace{-0.03cm}-\hspace{-0.02cm}0.7q^{\hspace{-0.02cm}-\hspace{-0.02cm}1})$.
Considering a sampling period $h=0.05$[s], the DT equivalent of $G_\textnormal{c}^*(p)$ has a parameter vector given by $\bm{\theta}_{\textnormal{d}}^* = [-\hspace{-0.02cm}1.069,\hspace{-0.01cm}0.546,\hspace{-0.02cm}-\hspace{-0.02cm}1.979,\hspace{-0.01cm}1.65,\hspace{-0.01cm}0.991,\hspace{-0.01cm}2.665,\hspace{-0.02cm}-\hspace{-0.02cm}2.241,\hspace{-0.02cm}-\hspace{-0.02cm}1.268]^{\hspace{-0.03cm}\top}\hspace{-0.04cm}$. The input is a ZOH-interpolated multisine with angular frequencies $\omega=1,1.9,2.1,18,22$ [\textnormal{rad}/\textnormal{s}], and the white noise $e(kh)$ filtered by $H^*(q)$ has variance equal to $6$, which gives a signal-to-noise ratio of approximately $26$ [dB]. If a DT model is sought, then one can use the RIV and SRIV estimators described in Section \ref{sec:RIV}. However, DT identification methods that offer more flexibility than these two are the ASRIV and ARIV methods, since these algorithms can adjust for the smoothness of the CT step response via the relative degree enforcement of the CT equivalent transfer function estimate.

Table \ref{table1} shows the mean square error (MSE) of each parameter when performing $500$ Monte Carlo runs of $N=10^4$ samples each, with varying noise realizations. While being competitive or marginally better in terms of the MSE of the denominator parameters, the ASRIV and ARIV methods can identify the numerator parameters with at least one order of magnitude less of MSE compared to their non-adapted counterparts. We note that this improvement has been achieved using only discrete-time tools; the ASRIV and ARIV methods are in fact equal at convergence to the DT equivalents of the estimates obtained from the SRIVC and RIVC methods, respectively.


\section{Conclusions}
\vspace{-0.02cm}
\label{conclusions}
In this paper, we proved that the CT equivalent of the RIV estimator shares the same limiting points with the RIVC estimator for relative degrees one and zero. Such relation fails for $m<n-1$ since the indirect approach generally does not produce an estimate with the desired model structure. We also introduced the Adapted RIV estimator, which provides the correct relative degree in its CT equivalent. Under a similar logic, it is shown that it is also possible to adapt the RIVC estimator such that its DT equivalent has relative degree greater than one by construction. These adapted estimators provide the missing link between the refined instrumental variable methods in the DT and CT domain.

\vspace{-0.1cm}
\bibliography{references.bib}

\begin{thebibliography}{10}

\bibitem{ljung1998system}
L.~Ljung, {\em System Identification: Theory for the User, \textnormal{2nd
  ed}}.
\newblock Prentice-Hall, 1999.

\bibitem{garnier2008book}
H.~Garnier and L.~Wang~(Eds.), {\em Identification of Continuous-time Models
  from Sampled Data}.
\newblock Springer, 2008.

\bibitem{gonzalez2018asymptotically}
R.~A. Gonz{\'a}lez, C.~R. Rojas, and J.~S. Welsh, ``An asymptotically optimal
  indirect approach to continuous-time system identification,'' in {\em 57th
  IEEE Conference on Decision and Control (CDC)}, pp.~638--643, 2018.

\bibitem{young1976some}
P.~C. Young, ``Some observations on instrumental variable methods of
  time-series analysis,'' {\em International Journal of Control}, vol.~23,
  no.~5, pp.~593--612, 1976.

\bibitem{garnier2015direct}
H.~Garnier, ``Direct continuous-time approaches to system identification.
  {O}verview and benefits for practical applications,'' {\em European Journal
  of Control}, vol.~24, pp.~50--62, 2015.

\bibitem{ljung2009experiments}
L.~Ljung, ``Experiments with identification of continuous time models,'' in
  {\em 15th IFAC Symposium on System Identification, \textnormal{Saint Malo,
  France}}, vol.~42, pp.~1175--1180, 2009.

\bibitem{young1980refined}
P.~C. Young and A.~J. Jakeman, ``Refined instrumental variable methods of
  recursive time-series analysis. {P}art {III}, {E}xtensions,'' {\em
  International Journal of Control}, vol.~31, no.~4, pp.~741--764, 1980.

\bibitem{garnier2021new}
H.~Garnier, M.~Gilson, H.~Muller, and F.~Chen, ``A new graphical user interface
  for the {CONTSID} toolbox for {M}atlab,'' {\em IFAC-PapersOnLine}, vol.~54,
  no.~7, pp.~397--402, 2021.

\bibitem{young2012recursive}
P.~C. Young, {\em Recursive Estimation and Time-Series Analysis: An
  Introduction for the Student and Practitioner, \textnormal{2nd Edition}}.
\newblock Springer, 2012.

\bibitem{pan2021continuous}
S.~Pan, Q.~C. Nguyen, V.~T. Nguyen, and J.~S. Welsh, ``Continuous-time system
  identification of a flexible cantilever beam,'' in {\em 2021 IEEE Conference
  on Control Technology and Applications (CCTA)}, pp.~868--873, 2021.

\bibitem{laurain2010refined}
V.~Laurain, M.~Gilson, R.~T{\'o}th, and H.~Garnier, ``{Refined instrumental
  variable methods for identification of LPV Box--Jenkins models},'' {\em
  Automatica}, vol.~46, no.~6, pp.~959--967, 2010.

\bibitem{ni2013refined}
B.~Ni, M.~Gilson, and H.~Garnier, ``{Refined instrumental variable method for
  Hammerstein--Wiener continuous-time model identification},'' {\em IET Control
  Theory \& Applications}, vol.~7, no.~9, pp.~1276--1286, 2013.

\bibitem{gonzalez2022refined}
R.~A. Gonz{\'a}lez, C.~R. Rojas, S.~Pan, and J.~S. Welsh, ``Refined
  instrumental variable methods for unstable continuous-time systems in
  closed-loop,'' {\em International Journal of Control}, pp.~1--15, 2022.

\bibitem{young2015refined}
P.~C. Young, ``Refined instrumental variable estimation: {M}aximum {L}ikelihood
  optimization of a unified {B}ox--{J}enkins model,'' {\em Automatica},
  vol.~52, pp.~35--46, 2015.

\bibitem{yuz2014sampled}
J.~I. Yuz and G.~C. Goodwin, {\em Sampled-Data Models for Linear and Nonlinear
  Systems}.
\newblock Springer, 2014.

\bibitem{kollar1996equivalence}
I.~Kollar, G.~Franklin, and R.~Pintelon, ``On the equivalence of z-domain and
  s-domain models in system identification,'' in {\em IEEE Instrumentation and
  Measurement Technology Conference}, vol.~1, pp.~14--19, 1996.

\bibitem{rudin1976principles}
W.~Rudin, {\em Principles of Mathematical Analysis, \textnormal{3rd Edition.}}
\newblock McGraw-Hill, 1976.

\bibitem{pan2020consistency}
S.~Pan, R.~A. Gonz{\'a}lez, J.~S. Welsh, and C.~R. Rojas, ``Consistency
  analysis of the {S}implified {R}efined {I}nstrumental {V}ariable method for
  {C}ontinuous-time systems,'' {\em Automatica}, vol.~113, \textnormal{Art.
  108767}, 2020.

\bibitem{gonzalez2022theoretical}
R.~A. Gonz{\'a}lez, C.~R. Rojas, S.~Pan, and J.~S. Welsh, ``Theoretical and
  practical aspects of the convergence of the {SRIVC} estimator for
  over-parameterized models,'' {\em Automatica}, vol.~142, \textnormal{Art.
  }110355, 2022.

\bibitem{astroem1984computer}
K.~J. {\AA}str{\"o}m and B.~Wittenmark, ``Computer {C}ontrolled {S}ystems:
  {T}heory and {D}esign,'' {\em Prentice-Hall}, 1984.

\bibitem{garnier2014advantages}
H.~Garnier and P.~C. Young, ``The advantages of directly identifying
  continuous-time transfer function models in practical applications,'' {\em
  International Journal of Control}, vol.~87, no.~7, pp.~1319--1338, 2014.

\bibitem{rao2002numerical}
G.~P. Rao and H.~Garnier, ``Numerical illustrations of the relevance of direct
  continuous-time model identification,'' in {\em 15th IFAC World Congress,
  \textnormal{Barcelona, Spain}}, vol.~35, pp.~133--138, 2002.

\end{thebibliography}
\end{document}